\documentclass[aps,prb,twocolumn]{revtex4} 

\usepackage{graphicx}
\usepackage{dcolumn}
\usepackage{bm}
\usepackage{amsmath}  

\newcommand{\comment}[1]{}

\begin{document}

\title{Information vs Thermodynamic Entropy}


\author{Phil Attard}
\affiliation{ {\tt phil.attard1@gmail.com} }


\begin{abstract}
The Shannon information is shown to be different
to the thermodynamic entropy,
and indifferent to the Second Law of Thermodynamics.
\end{abstract}


\maketitle

%
%

\section{The measure of a name} \label{Sec:Name}
\sectionmark{\S\ref{Sec:Name} The measure of a name}

In 1948 Claude E. Shannon published a paper
on the mathematical theory of communications
that quickly became a classic.
It established the field of information theory,
which is vital to disciplines
such as genomics, cryptology,
computer science, network engineering, signal processing,
data storage,  quantum informatics, to name a few.
Shannon proposed what he initially called
a logarithmic measure of information (Shannon 1948),
\begin{equation} \label{Eq:H[wp]}
H[\wp_i] = -k_\mathrm{B} \sum_i \wp_i \ln \wp_i ,
\end{equation}
where $\wp_i$, the probability of the microstate $i$,
is normalized to unity.
Nowadays this is usually called the entropy, or the information entropy,
but for reasons that will quickly become clear
I prefer to call it the Shannon information.
It is not essential to use for the prefactor
Boltzmann's constant $k_\mathrm{B}=1.38\times10^{-23}$J/K,
nor to use base $e=2.718\ldots$ for the logarithm.

Shannon gave certain plausible axioms
that any measure of information should obey,
but conceded that
`[t]he real justification of these
definitions, however, will reside in their implications'
(Shannon 1948 \S6). 
Notably he did not derive this functional form
from thermodynamical or statistical mechanical arguments,
but nevertheless stated
`We shall call $H = -  \sum_i p_i \ln p_i $
the entropy of the set of probabilities $p_1,\ldots, p_n$'
(Shannon 1948 \S6). 
Shannon also called it
a `measure of choice or information' or `the uncertainty'
(Shannon 1948 \S6). 

Shannon appears to have been persuaded to call this the entropy
by von Neumann (Tribus and McIrvine 1971).
von Neumann had already given an equivalent trace form for the entropy
in his presentation of quantum statistical mechanics (von Neumann 1932).
But even before that, as mentioned by Shannon (1948),
the same functional was given by Boltzmann,
who called it the $H$-function (Boltzmann 1866).
And it was also given by Gibbs
in his formulation of thermodynamics,
who called it the average of the index of probability
(Gibbs 1902).
Although these two may be regarded as the respective founders
of statistical mechanics and of modern thermodynamics,
neither called this functional the entropy.

Since von Neumann's efforts, this functional
has been routinely  called the entropy
in textbooks in statistical mechanics
and in the physical sciences.
Perhaps the most notable proponent 
has been Jaynes (1957, 2003),
whose so-called maxent approach asserts
that this functional is maximal
with respect to the probability distribution
constrained by explicit knowledge about the system.
This relies upon the Second Law of Thermodynamics,
which would be fine if  this functional
were indeed the thermodynamic entropy.
Shannon (1948)
does not refer to the  Second Law of Thermodynamics,
but does observe that the functional is a maximum
for a uniform probability distribution,
which is `intuitively the most uncertain situation'
(Shannon 1948 \S6). 

Common books and papers
are based upon the maximisation of this functional,
and the assertion that in physical systems it increases over time.
These would be credible if the functional
were indeed the thermodynamic entropy
that is the subject of the Second Law of Thermodynamics,
but not otherwise.

Shannon (1948) makes an important point about the information
contained in a set of macrostates, labeled $\alpha$,
each of which is a set of microstates, labeled $i$.
(A microstate is the smallest possible division of the system.)
The  probability of a macrostate is $\wp_\alpha = \sum_{i\in \alpha} \wp_i$,
and the conditional probability is $\wp_{i|\alpha} = \wp_i/\wp_\alpha$,
${i\in \alpha}$.
In this case the total uncertainty is the sum of the uncertainty
that the system is in a macrostate,
plus the weighted sum of the uncertainties
due to it being in a microstate within a macrostate
(Shannon  1948),
\begin{equation} \label{Eq:compo}
H[\wp_\alpha;\wp_{i|\alpha}]
=
H[\wp_\alpha]
+ \sum_\alpha \wp_\alpha H[\wp_{i|\alpha}] .
\end{equation}
(Here $H[\wp_\alpha]$ is the mathematical function of its argument
rather than the physical total uncertainty of the system.)
It is straightforward to show from this
that the amount of information is the same whether calculated
from macrostates or from microstates
\begin{eqnarray}
\lefteqn{
H[\wp_\alpha;\wp_{i|\alpha}]
} \nonumber \\
& = &
-k_\mathrm{B}  \sum_\alpha \wp_\alpha \ln \wp_\alpha
-k_\mathrm{B} \sum_\alpha \wp_\alpha
\sum_{i\in\alpha} \wp_{i|\alpha} \ln \wp_{i|\alpha}
\nonumber \\ & = &
-k_\mathrm{B}  \sum_\alpha \wp_\alpha \ln \wp_\alpha
-k_\mathrm{B} \sum_\alpha \wp_\alpha
\sum_{i\in\alpha} \frac{\wp_i}{\wp_\alpha} \ln \wp_{i}
\nonumber \\ && \mbox{ }
+k_\mathrm{B} \sum_\alpha \wp_\alpha
\sum_{i\in\alpha} \frac{\wp_i}{\wp_\alpha} \ln \wp_\alpha
\nonumber \\ & = &
-k_\mathrm{B}  \sum_i \wp_i \ln \wp_i
\nonumber \\ & = &
H[\wp_{i}].
\end{eqnarray}
The distinction between macrostates and microstates is important,
and it is essential that the respective formula be used in each case.
The total information is not correctly given
using only the macrostate probability,
$H[\wp_{\alpha}] \ne H[\wp_{i}]$.
Many fail to avert to this distinction.

The law of composition, Eq.~(\ref{Eq:compo}),
says that the information or uncertainty is independent of how microstates
are grouped into macrostates provided
that the additional uncertainty of the grouping is accounted for.
It is this axiom that leads to the definition of the information
as Shannon invokes it explicitly
in the mathematical derivation of his functional
(Shannon 1948 Appendix 2). 
This point is important for the discussion of the similarities
and differences with the thermodynamic entropy.

\section{Informative thermodynamics} \label{Sec:TD}
\sectionmark{\S\ref{Sec:TD} Informative thermodynamics}

Statistical thermodynamics requires more information
than information theory.
The formulation of the Shannon information just summarized
only requires the microstate probabilities $\wp_i$
and the macrostate probabilities $\wp_\alpha = \sum_{i\in\alpha} \wp_i$.
These are of course  normalized to unity.

The formulation of statistical thermodynamics
requires the microstate weights $w_i$,
which are real and non-negative,
the macrostate weights $W_\alpha = \sum_{i\in\alpha} w_i$,
and the total weight
\begin{equation}
W = \sum_i w_i = \sum_\alpha W_\alpha.
\end{equation}
From these the microstate probability,
$\wp_i = w_i/W$,
and the macrostate probability, $\wp_\alpha = W_\alpha/W$, follow.
The point is that one extra piece of information
is required for statistical thermodynamics, namely the total weight.

In general the weights are either measured experimentally,
or else calculated from statistical mechanics
according to the known dynamical laws.
Here we need not enquire too deeply of them,
except to say that in the simplest classical case,
the microstates are the points in classical phase space,
and for an isolated system this has real, non-negative, uniform weight
(Attard 2002, 2012a, 2023).
In the quantum case the microstates can be taken to be
the points in the Hilbert space of unnormalized wave functions
(equivalently,  the eigenstates of a complete operator),
and again in the isolated case
the weights are real, non-negative, and uniform (Attard 2023).
(The use of weight in these cases is a minor generalization
of Boltzmann's original prescription
of the number of molecular configurations.)

The microstate weights of an open system
(ie.\ one that can exchange with an environment, reservoir, or heat bath)
are not uniform, but they can be calculated
from the usual conservation laws and thermodynamic definitions
(Attard 2002).
Again these details do not concern us here.

The detail that does concern us is that the weight has physical meaning.
In particular the thermodynamic entropy of the system
is the logarithm of the total weight
\begin{equation}
S = k_\mathrm{B} \ln W .
\end{equation}
One can similarly define the entropy of a microstate
and of a macrostate,
\begin{equation}
S_i = k_\mathrm{B} \ln w_i ,
\mbox{ and }
S_\alpha = k_\mathrm{B} \ln W_\alpha.
\end{equation}
With these the probabilities are simply proportional to
the exponentials of the entropies,
\begin{equation}
\wp_i = \frac{1}{W} e^{ S_i /k_\mathrm{B} } ,
\mbox{ and }
\wp_\alpha = \frac{1}{W} e^{ S_\alpha /k_\mathrm{B} }.
\end{equation}

It ought to be clear that the entropy
that goes into the Second Law of Thermodynamics
is the thermodynamic one.
In fact the Second Law follows directly from these expressions
since the optimum state is the most likely state,
which is the state of maximum entropy.
It is also obvious that the likely direction for transitions between states
is that of increasing state probability,
which is that of increasing entropy (Attard 2012a).
In other words these expressions explain the origin and the validity
of the  Second Law of Thermodynamics,
and they make explicit the fact that the thermodynamic entropy
is the logarithm of the weight.

It also ought to be clear that there is a one-to-one relationship
between the entropy of a state and the probability of a state.
Hence it is nonsensical to attempt to determine the probabilities
by maximizing the thermodynamic entropies,
which in any case are fixed by the physical characteristics of the system.

\section{Incomplete entropy} \label{Sec:IneTD}
\sectionmark{\S\ref{Sec:IneTD} Incomplete entropy}

The Shannon information of a system
is not equal to the thermodynamic entropy of the system.
This is easily shown from the definitions,
\begin{eqnarray}
H[\wp_i] & = &
-k_\mathrm{B} \sum_i \wp_i \ln \wp_i
\nonumber \\ & = &
-k_\mathrm{B} \sum_i \wp_i \ln \frac{e^{S_i/k_\mathrm{B}}}{W}
\nonumber \\ & = &
k_\mathrm{B} \ln W -  \sum_i \wp_i S_i
\nonumber \\ & = &
S - \left\langle S^\mathrm{micro} \right\rangle .
\end{eqnarray}
We see that the Shannon information of the system
is equal to the thermodynamic entropy of the system
less the average microstate entropy.
In general the latter cannot be set to zero because
the weights are non-uniform and are fixed by the physical characteristics
of the thermodynamic system.
As will be seen next by explicit example,
the average microstate entropy is extensive with system size
and so again it cannot be neglected.

As mentioned above it was von Neumann
who was primarily responsible for convincing Shannon
to identify  the Shannon information with the thermodynamic entropy
(Tribus and McIrvine 1971).
von Neumann (1932)
had already given this functional form
in his density matrix formulation of quantum statistical mechanics,
calling it the entropy.
It therefore seems fitting to choose a concrete example
from quantum statistical mechanics to demonstrate explicitly
that the Shannon information is not equal to the thermodynamic entropy.

Consider a system of non-interacting bosons
with energy states ${\bf a}$
of energy $\varepsilon_{\bf a}$.
For an open system
a particular set of state occupancies,
$\underline N = \{\ldots, N_{\bf a}, \ldots \}$,
has probability
\begin{equation}
\rho(\underline N)
=
\frac{1}{\Xi^+(z,V,T)}
\prod_{\bf a} z^{N_{\bf a}} e^{-\beta N_{\bf a} \varepsilon_{\bf a} } ,
\end{equation}
where
$\beta=1/k_\mathrm{B}T$ is the inverse temperature,
$z=e^{\beta \mu}$ is the fugacity,
$\mu$ being the chemical potential,
and $\Xi^+$ is the grand partition function,
which is equivalent to the total weight $W$.
This is also called the density matrix,
and the Shannon information is given by its von Neumann trace,
\begin{eqnarray}
H[\rho] & = &
-k_\mathrm{B}\mbox{TR } \rho \ln \rho
\nonumber \\ & = &
-k_\mathrm{B}\prod_{\bf a} \sum_{N_{\bf a}=0}^\infty
\rho(\underline N)
\ln
\frac{ \prod_{\bf a} z^{N_{\bf a}} e^{-\beta N_{\bf a} \varepsilon_{\bf a}}
}{ \Xi^+(z,V,T) }
\nonumber \\ & = &
k_\mathrm{B}\ln \Xi^+(z,V,T)
\nonumber \\ && \mbox{ }
-
k_\mathrm{B}\prod_{\bf a} \sum_{N_{\bf a}=0}^\infty
\rho(\underline N)
\ln z^{N_{\bf a}} e^{-\beta N_{\bf a} \varepsilon_{\bf a}}
\nonumber \\ & = &
k_\mathrm{B}\ln \Xi^+(z,V,T)
- \frac{\mu}{T} \langle N \rangle
+ \frac{1}{T} \langle E \rangle
\nonumber \\ & = &
S^\mathrm{tot}(z,V,T) - S^\mathrm{r}(z,V,T)
\nonumber \\ & = &
S^\mathrm{s}(z,V,T) .
\end{eqnarray}
As usual in statistical mechanics,
the total entropy of the total system
is the logarithm of the partition function,
which  is the total weight.
This is the sum 
of the subsystem-dependent part of the reservoir entropy,
$S^\mathrm{r} = [\mu  N  -  E ]/T$,
and the subsystem entropy, $S^\mathrm{s}$ (Attard 2002 \S3.2).
(For the present purposes, we don't need to distinguish between
the total unconstrained entropy,
the maximal constrained entropy,
and the average constrained entropy (Attard 2002).)
We see from this that
the von Neumann trace gives only the subsystem contribution
to the total entropy.
It neglects the reservoir
(ie.\ environment or heat and particle bath) contribution.

This concrete example shows the folly of attempting to maximize
the Shannon information to obtain the optimum state.
The penultimate equality shows explicitly
that the Shannon information
is only part of the total thermodynamic entropy.
The Second Law of Thermodynamics
provides no prescription for the behavior of part of the entropy.

\section{Information confusion}

The discussion of Eq.~(\ref{Eq:compo}) above
raises a subtle difference between mathematics and the physical sciences.
In mathematics the notation $H[\wp]$ means to take the argument
of the function and to use it on the right hand side
of the original definition of that function
in place of the argument used in its definition.
However in the physical sciences, different symbols are used
to denote different physical quantities,
and so in the present context $H$
is meant to be the total information of the system,
not just part of the information.
The problem is that on the right hand side
of Eq.~(\ref{Eq:compo}),
$H(\wp_\alpha)$ is used in the mathematical sense
(ie.\ the mathematical function, Eq.~(\ref{Eq:H[wp]})
with $\wp_\alpha$ on the right hand side),
not in the physical sense (ie.\ the total information of the system).
It is the left hand side of Eq.~(\ref{Eq:compo})
that is the total information of the system.
In the physical sciences the argument  of a function
can only be replaced by a quantity of the same physical type.
In the present case the microstate probability $\wp_i$
is qualitatively different to the macrostate probability $\wp_\alpha$,
at least from the point of view of information,
but both are used as arguments in the mathematical function $H[\wp]$.
These different conventions can cause confusion,
and they underscore the importance of stating explicitly that
$H[\wp_i]$ is the total information if and only if $\wp_i$
is the microstate probability.

Attard (2012b)
correctly showed that the Shannon information
is different to the thermodynamic entropy,
but incorrectly assumed that the two \emph{should} be the same.
Attard's (2012b) analysis,
which purported to prove
that Shannon's derivation of the information functional is erroneous,
assumed that the function of the macrostate probability,
$H[p_i]$,
which appeared on the right hand side of the penultimate equation
of Appendix~2 of Shannon (1948),
was the physical total information,
when in fact Shannon was using it in the mathematical sense,
which has the physical interpretation
of the partial information due to macrostate probabilities.
In fact Shannon's (1948 Appendix~2) derivation of Eq.~(\ref{Eq:H[wp]})
is sound provided that $H[p_i]$
is interpreted mathematically rather than physically,
and that the Shannon information
is not equated to the thermodynamic entropy.

\section{Conclusion} \label{Sec:Concl2}
\sectionmark{\S\ref{Sec:Concl2} Conclusion}

The difference between the Shannon information
and the thermodynamic entropy
is due to  the principle of composition,
Eq.~(\ref{Eq:compo}).
This says that for macrostates composed of microstates,
the uncertainty is the sum of the uncertainty
that the system is in a macrostate,
plus the weighted sum of the uncertainties
due to it being in a microstate within a macrostate.
This is a  plausible requirement for information,
and it is difficult to see how one would proceed without it.

The principle of composition means that the functional for
the Shannon information differs
for the microstate probability, Eq.~(\ref{Eq:H[wp]}),
from that for the macrostate probability, Eq.~(\ref{Eq:compo}).
In contrast, the formula for the thermodynamic entropy
is the same function of weight
for microstates as for macrostates,
$S = k_\mathrm{B} \ln \sum_i w_i
= k_\mathrm{B} \ln \sum_\alpha W_\alpha $.
Consequently, the thermodynamic entropy
does not obey the principle of composition.

The fact that the Shannon information
is different to the thermodynamic entropy
does not mean that one is right and the other is wrong.
Rather it means that they have different uses and applications.
It also means that the properties of one
cannot be assumed to hold for the other.

In particular, the Second Law of Thermodynamics
applies to the thermodynamic entropy,
but not to the Shannon information.
The fact that the uniform microstate probability distribution
maximises the Shannon information
may or may not be useful in informatic and communications applications.
Whilst I am  dubious that this provides
a variational principle of any universality
for these and related fields,
I'm willing to consider that specific applications may be exceptions.

In practice there is no drive toward uniformity
in the progress of a signal or message over time.
Nor are most messages dominated by uniformly random sequences of symbols.
For example,
the uncertainty contributed by this paper
is more or less the same at the beginning as at the end,
and it is not nearly as large as it could be.


\section*{References}


\begin{list}{}{\itemindent=-0.5cm \parsep=.5mm \itemsep=.5mm}

\item 
Attard P 2002
\emph{Thermodynamics and Statistical Mechanics:
Equilibrium by Entropy Maximisation}
(London: Academic)

\item 
Attard  P 2012a
\emph{Non-equilibrium thermodynamics and statistical mechanics:
Foundations and applications}
(Oxford: Oxford University Press)

\item
Attard  P 2012b
Is the Information Entropy
the Same as the Statistical Mechanical Entropy?
arXiv:1209.5500v1.

\item 
Attard P 2023
\emph{Entropy beyond the Second Law:
Thermodynamics and statistical mechanics for equilibrium, non-equilibrium,
classical, and quantum systems}
(Bristol: IOP Publishing, 2nd edition)

\item 
Boltzmann L 1866
\"Uber die mechanische bedeutung des zweiten hauptsatzes
der w\"armetheorie
\emph{Wiener Berichte} {\bf  53} 195 

\item 
Gibbs J W 1902
\emph{Elementary Principles in Statistical Mechanics Developed with
Special Reference to the Rational Foundation of Thermodynamics}
(New Haven, CT: Yale Univ.\ Press)

\item 
Jaynes E T 1957
Information theory and statistical mechanics
\emph{Phys.\ Rev.}\ {\bf 106} 620  and {\bf 108} 171

\item 
Jaynes E T  2003
\emph{Probability Theory: The Logic of Science}
ed G L Bretthorst,  (Cambridge: Cambridge University Press)

\item 
von Neumann J 1932
\emph{Mathematische Grundlagen der Quantenmechanik}
(Berlin: Springer).
von Neumann J 2018
\emph{Mathematical Foundations of Quantum Mechanics}
R T Beyer (Trans.),  N A Wheeler (Ed.)
(Princeton: Princeton University Press).

\item 
Shannon C E 1948
A mathematical theory of communication
\emph{Bell Syst.\ Tech.\ J.}\ {\bf 27} 379, 623.
{\tt www.alcatel-lucent.com/bstj/vol27-1948
/articles/bstj27-3-379.pdf}


\item 
Tribus M and McIrvine E C 1971
Energy and information science
\emph{Sci.\ Am.}\ {\bf 224} 179

\end{list}



%
%

\end{document}